\documentclass[preprint,number, authoryear,12pt]{article}
\usepackage[margin=3cm]{geometry}

\title{ \bf Orthonormal Frame and SO(3) Kaluza-Klein Dyon  \rm}

\author{\bf  Hsiang-Lin Lou\thanks{~Email: {sllou@thu.edu.tw}}
          and   Pei-Hsuan  Li\rm \thanks{~Email: {ludmila.phys@gmail.com}}\\
       Deparment of Physics, Tunghai University\\
       Taichung, 407, Taiwan}
\date{\today}

\begin{document}
\maketitle
\vspace{3cm}
\begin{abstract}
 In previous paper, we present an SO(3) Wu-Yang-like Kaluza-Klein dyon solution
  satisfies the Einstein equation in
 the seven-dimensional
spacetimes. In this note, we will show an alternative approach  using an orthonormal frame,
 the Cartan's structure equations, and  calculating the affine spin connection one-form,
curvature tensor and Ricci tensor.  The results from these two different methods are coincident.

\end{abstract}
\vspace{1cm} \noindent PACS numbers: 04.50.Cd

\vspace{6cm}
%%%%%%%%%%%%%%%%%%%%%%%%%%%%%%%%%%%%%%%%%
\section{Introduction}
In previous paper,[1] we have shown that a new SO(3) Wu-Yang-like [2]
seven-dimensional  Kaluza-Klein(KK)  dyon solution
 satisfies the Einstein equation, by calculating the Christoffel symbols and  the  Ricci tensor.
  In this note, we will show  an alternative approach [3]  using an orthonormal frame,
 the Cartan's structure equations and  calculating the affine spin connection one-form,
curvature tensor and Ricci tensor.

Now  consider the Kaluza-Klein theory [4] in  a $(4+N)$-manifold $M$ with
a metric $\bar{g}_{AB} (x)$ on $M$ in local coordinates $\bar{x}^{A}$. The line element is
  \begin{eqnarray}
  d \bar{s}^2 &=&\bar{g}_{AB}d\bar{x}^A d\bar{x}^B  \\
  &=&g_{\mu \nu} (x) dx^{\mu} dx^{\nu}    +
  \gamma_{mn}(y) (dy^m + B_{\mu}^m dx^{\mu} )(dy^n + B_{\nu}^n dx^{\nu} ),
  \end{eqnarray}
 where $x$ parametrizes   four-dimensional spacetimes, $y$ parametrizes extra dimensions.
 We use $A,B, C...$ indices to represent the total spacetimes; $\mu , \nu , \rho...$ to represent the four-dimensional
 spacetimes; $m, n, l...$ to represent the extra dimensions. $g_{\mu \nu}$ is only a function
 of $x$, and $\gamma_{mn}$ is only a function of $y$.

 The  ans$\ddot{a}$tz [1] of the  Kaluza-Klein dyonic metric admitting $SO(3)$
Killing vectors is
\begin{eqnarray}
d\bar{s}^2 =& -& e^{2\Psi} dt^{2} \nonumber \\
 &+& e^{2\Lambda} dr^{2} +r^2 d\theta^2 +r^2 \sin^{2}{\theta} d\phi^2
\nonumber \\
&+& (dR + B_{\mu}^{5} dx^{\mu} )^2
+ R^2 (d\Theta  + B_{\mu}^{6} dx^{\mu} )^2
+ R^2 \sin^{2}\Theta (d\Phi  + B_{\mu}^{7} dx^{\mu} )^2 .
\end{eqnarray}
$r,\theta , \phi $ are  three coordinates of the  ordinary   three-dimensional spherical coordinate system,
$(r,\theta ,\phi )=(\bar{x}^1 ,\bar{x}^2 , \bar{x}^3 )=(x^1 ,x^2 , x^3 )$.
 $R, \Theta , \Phi$
are another three coordinates of  the spherical coordinate system in the extra dimensions,
 $(R, \Theta ,\Phi ) =(\bar{x}^5 ,\bar{x}^6 , \bar{x}^7 )=(y^{5} ,y^6 ,y^7 )$.
$\Psi$ and $\Lambda$ are two functions of $r$.
 We have $g_{00}=-e^{2\Psi}$, $g_{11}=e^{2\Lambda}$, $g_{22} =r^2$, $g_{33}=r^2 \sin^2 \theta$, $\gamma_{55}=1$,
 $\gamma_{66}=R^2$, $\gamma_{77}=R^2 \sin^2 \Theta$.

$B_{\mu}^m $ cannot be identified as the Yang-Mills field. To extract the true Yang-Mills field,
one has to introduce the Killing vectors
\begin{equation}
L_a \equiv - i \zeta_{a}^{m} \partial_{m},
 \end{equation}
which generate an $SO(3)$ algebra,
\begin{equation}
[L_{a} , L_{b} ]=i \epsilon_{abc} L_{c},
 \end{equation}
Inserting $L_{a}$ of (4)
into equation (5), one gets the Killing's equation
\begin{equation}
\zeta_{a}^m \partial_{m} \zeta_{b}^n -
\zeta_{b}^m \partial_{m} \zeta_{a}^n
=- \epsilon_{abc} \zeta_{c}^n .
 \end{equation}
 With these Killing vectors, one can define
 \begin{equation}
B_{\mu}^m =\zeta_{a}^{m} A_{\mu}^{a},
 \end{equation}
where $A_{\mu}^{a}$ is the true Yang-Mills field and
$\zeta_{a}^{m}$ is only a function of $y$.
Defining
\begin{eqnarray}
\widetilde{\mathcal{F}}_{\mu \nu}^{m} &\equiv&
\partial_{\mu} B_{\nu}^{m} + B_{\nu}^{l} \partial_{l} B_{\mu}^{m}
-(\mu \leftrightarrow \nu ) \\
&=& \zeta_{a}^{m} F_{\mu \nu}^{a},
\end{eqnarray}
 where
 $F_{\mu \nu}^{a}$  is the true field strength tensor of the
Yang-Mills field,
\begin{equation}
F_{\mu \nu}^{a} = \partial_{\mu} A_{\nu}^{a}- \partial_{\nu} A_{\mu}^{a}
+\epsilon_{abc} A_{\mu}^{b} A_{\nu}^{c}.
\end{equation}
It can be checked that the components of $\zeta_{a}^{m}$,
\begin{eqnarray}
\zeta_{a}^{5} &=& 0, \\
\zeta_{a}^{6} &=& \hat{\Phi}_{a}, \\
\zeta_{a}^{7} &=& -{1\over \sin{\Theta} } \; \hat{\Theta}_{a},
\end{eqnarray}
satisfy the Killing equation (6).
The gauge-field components of  the Wu-Yang-like KK dyon   are
\begin{eqnarray}
A_{0}^{a} &=& {1\over r} \hat{r}^{a}, \;\; \; \;\;\; \; \;
A_{1}^{a}  = 0, \\
A_{2}^{a} &=& - \hat{\phi}^{a}, \; \;\; \; \; \;\;
A_{3}^{a}  =  \sin\theta  \; \hat{\theta}^{a} .
\end{eqnarray}
$A_{1}^{a}$,$A_{2}^{a}$,$A_{3}^{a}$ are just the spherical coordinate representation of the Wu-Yang
monopole field in the ordinary gauge theory of four-dimensional spacetimes. The electric field of the
KK  dyon is
\begin{equation}
F_{01}^{a} = {1\over r^2} \;  \hat{r}^{a} ,
\end{equation}
while the magnetic field  is
\begin{equation}
F_{23}^{a} = - \sin \theta  \; \hat{r}^{a}.
\end{equation}
The fields $B_{\mu}^{m}$ in (7) can be rewritten as
\begin{eqnarray}
B_{\mu}^{5} &=& \; 0,    \; \; \; \; \; \; \; \; \; \; \; \; \; \;  \; \; \; \; \; \;
B_{1}^{m} =\; \; 0,\; \; \; \; \; \; \; \; \; \; \; \;  \; \; \; \; \;  \\
B_{0}^{6}&=& {1\over r} \; \hat{r}  \cdot  \hat{\Phi}, \; \; \; \; \; \; \; \; \; \; \; \; \; \;
B_{0}^{7} = - { 1\over  {r\sin\Theta}} \; \hat{r} \cdot  \hat{\Theta}, \\
B_{2}^{6}&=&   - \hat{\phi}  \cdot  \hat{\Phi}, \; \; \; \; \; \; \; \; \; \; \; \;\; \;
B_{2}^{7} = \; \;\;  { 1\over  {\sin\Theta}} \; \hat{\phi} \cdot  \hat{\Theta}, \\
B_{3}^{6}&=&  \; \sin\theta \;  \hat{\theta}  \cdot  \hat{\Phi}, \; \; \; \; \; \; \; \; \;
B_{3}^{7} = - {\sin\theta \over  {\sin\Theta}} \; \hat{\theta} \cdot  \hat{\Theta}.
\end{eqnarray}
The nonzero components of $\widetilde{\mathcal{F}}_{\mu \nu}^{m}$ are
 \begin{eqnarray}
 \widetilde{\mathcal{F}}_{01}^{6} &=& \; \; {1\over r^2}\;  \hat{r} \cdot  \hat{\Phi},
\;\; \; \; \; \; \; \; \; \; \; \; \; \; \;
\widetilde{\mathcal{F}}_{01}^{7} = {-1 \over r^2\sin\Theta}\;  \hat{r} \cdot  \hat{\Theta},
\; \; \; \; \; \; \; \\
\widetilde{\mathcal{F}}_{23}^{6} &=&  {- \sin\theta}\;  \hat{r} \cdot  \hat{\Phi},
\;\; \; \; \; \; \; \; \; \; \;
\widetilde{\mathcal{F}}_{23}^{7} = {\sin\theta \over \sin\Theta}\;  \hat{r} \cdot  \hat{\Theta}.
\; \; \; \; \; \; \;
\end{eqnarray}

\section{Orthonormal Frame}

We now decompose the metric into vielbeins as
\begin{equation}
\bar{g}_{AB} =\eta_{\bar{a} \bar{b}}\;  e^{\bar{a}}_{\; A} e^{\bar{b}}_{\; B},
\; \; \; \; \; \; \; \; \bar{a},\bar{b}= 0,1,2,3,5,6,7,
\end{equation}
where the $\eta_{\bar{a} \bar{b}}$ is a seven-dimensional flat Minkowski space metric,
\begin{equation}
\eta_{\bar{a} \bar{b}} = diag(-1, 1,1,1,1,1,1,)
\end{equation}
The inverse of $e^{\bar{a}}_{\; A} $ is defined by
\begin{equation}
E_{\bar{a}}^{\; A} =\eta_{\bar{a} \bar{b}}\; \bar{g}^{AB}\;  e^{\bar{b}}_{\; B}
\end{equation}
which obeys
\begin{equation}
E_{\bar{a}}^{\; A}\;  e^{\bar{b}}_{\; A}= \delta_{\bar{a}}^{\bar{b}} ,
\; \; \; \; \; \; \; \; \; \; \; \; \; \; \;
\eta^{\bar{a} \bar{b}}\; E_{\bar{a}}^{\; A}\; E_{\bar{b}}^{\; B}=\bar{g}^{AB}.
\end{equation}
$e^{\bar{a}}_{\; A}$ is the matrix which tranforms the coordinate basis $d\bar{x}^{A}$
of the cotangent space $T^{*}_{x} (M) $ to an orthonormal basis of the same space $T^{*}_{x} (M) $,
\begin{equation}
e^{\bar{a}} =\; e^{\bar{a}}_{\; A} \; d{\bar{x}}^{A}.
\end{equation}
The vielbein basis of the SO(3) KK dyonic metric can be written as
\begin{equation}
e^0 =e^{\Psi} dt , \; \; \; \; \; \;   e^{1} =e^{\Lambda}\;  dr, \; \; \; \; \; \;
e^2 = r \; d\theta , \; \; \; \; \; \; e^3 = r sin\theta \; d\phi,
\end{equation}
\begin{equation}
e^5 =dR , \; \; \; \; \; \; \; \; \;   e^{6} =R(d\Theta +B^{6}_{\mu} dx^{\mu} ), \; \; \; \; \; \;
\; \; \; e^7 = R sin\Theta (d\Phi +B^{7}_{\mu} dx^{\mu} ).
\end{equation}

The affine spin connection one-form $\omega^{\bar{a}}_{\; \; \bar{b}}$
are introduced by
\begin{equation}
de^{\bar{a}} +\omega^{\bar{a}}_{\; \; \bar{b}} \wedge e^{\bar{b}} =\; 0
\end{equation}
and the metricity condition
\begin{equation}
\omega_{\bar{a} \bar{b}} = -\omega_{\bar{b} \bar{a}}.
\end{equation}

The cuvature 2-form is defined as
\begin{equation}
{{R}}^{\bar{a}}_{\; \; \bar{ b}} =d{\omega}^{\bar{a}}_{\; \; \bar{b}}
+{\omega}^{\bar{a}}_{\; \; \bar{c}}\wedge {\omega}^{\bar{c}}_{\; \; \bar{b}}
={\bar{R}}^{\bar{a}}_{\; \; {\bar{b}}{ \bar{c}}{ \bar{d}}}\;  e^{\bar{c}} \wedge e^{\bar{d}}.
\end{equation}
Equations (31) and (33) are called Cartan's structure equations. The components
of the curvature tensor  have the relations,
\begin{equation}
{\bar{R}}_{\bar{a} \bar{b} \bar{c} \bar{d}} =-{\bar{R}}_{\bar{b} \bar{a} \bar{c} \bar{d}}
=-{\bar{R}}_{\bar{a} \bar{b} \bar{d} \bar{c}}= {\bar{R}}_{\bar{c} \bar{d} \bar{a} \bar{b}}.
\end{equation}
and satisfy the Bianchi identity,
\begin{equation}
{\bar{R}}_{\bar{a} \bar{b} \bar{c} \bar{d}} + {\bar{R}}_{\bar{a} \bar{c} \bar{d} \bar{b}}
+ {\bar{R}}_{\bar{a} \bar{d} \bar{b} \bar{c}}= 0.
\end{equation}

The components of the affine spin connection one-form $\omega^{\bar{a}}_{\; \; \bar{b}}$ of the $SO(3)$
KK dyon can be written  more explicitely as
\begin{eqnarray}
\omega^0_{\; \; 1}& = &\Psi^{'} e^{-\Lambda} \; e^0 + {R\over 2r^2} e^{-\Psi -\Lambda} (\hat{r} \cdot \hat{\Phi})\; e^6
-{R\over 2r^2} e^{-\Psi -\Lambda} (\hat{r} \cdot \hat{\Theta})\;  e^7 , \; \; \; \; \; \; \; \; \;
 \; \; \; \; \; \; \; \; \;
  \; \; \; \; \; \; \;  \\
\omega^0_{\; \; 2} &=& 0,
 \; \; \; \; \; \; \; \; \; \; \; \; \; \; \; \; \; \; \; \; \; \; \omega^0_{\; \; 3} =0,
 \; \; \; \; \; \; \; \; \; \; \; \; \; \; \; \; \; \; \; \; \; \; \omega^0_{\; \; 5} =0,
 \; \;\; \; \; \; \; \; \; \; \; \; \; \; \; \; \\
 \omega^0_{\; \; 6}& = &  {R\over 2r^2} e^{-\Psi -\Lambda} (\hat{r} \cdot \hat{\Phi})\; e^1,
 \; \; \; \; \; \; \; \\
  \omega^0_{\; \; 7}  &= &-  {R\over 2r^2} e^{-\Psi -\Lambda} (\hat{r} \cdot \hat{\Theta})\; e^1, \\
   \omega^1_{\; \; 2}  &= &-  {1\over r} e^{ -\Lambda}\;  e^2, \\
  \omega^1_{\; \; 3}  &= &-  {1\over r} e^{ -\Lambda} \; e^3,
  \; \; \; \; \; \; \; \; \; \; \; \; \; \; \; \; \; \; \; \; \; \;  \; \; \; \; \; \; \;
  \omega^1_{\; \; 5} = 0, \\
 \omega^1_{\; \; 6}& = & \; \; {R\over 2r^2} e^{-\Psi -\Lambda} (\hat{r} \cdot \hat{\Phi})\; e^0 \\
 \omega^1_{\; \; 7}& = & -{R\over 2r^2} e^{-\Psi -\Lambda} (\hat{r} \cdot \hat{\Theta})\; e^0 \\
 \omega^2_{\; \; 3}& = & -{1\over r} cot\theta \; e^3
 + {R\over 2r^2} \;  (\hat{r} \cdot \hat{\Phi})\; e^6
-{R\over 2r^2} \;  (\hat{r} \cdot \hat{\Theta})\;  e^7,
 \; \; \; \; \; \; \; \; \; \; \;\omega^2_{\; \; 5} = 0, \\
 \omega^2_{\; \; 6}& = & \; \; {R\over 2r^2} (\hat{r} \cdot \hat{\Phi})\; e^3 ,\\
 \omega^2_{\; \; 7}& = & -{R\over 2r^2} (\hat{r} \cdot \hat{\Theta})\; e^3 ,
  \; \; \; \; \; \; \; \; \; \; \; \; \; \; \; \; \; \; \; \;  \omega^3_{\; \; 5} = \; \; 0,\\
 \omega^3_{\; \; 6}& = & -{R\over 2r^2}  (\hat{r} \cdot \hat{\Phi})\; e^2 ,
 \; \; \; \; \; \; \; \; \; \; \; \; \; \; \; \; \; \; \; \;
 \omega^3_{\; \; 7}  =  \; \;  {R\over 2r^2} \;  (\hat{r} \cdot \hat{\Theta})\; e^2 ,\\
 \omega^5_{\; \; 6}& = & -{1\over R}\;  e^6,
  \; \; \; \; \; \; \; \; \; \; \; \; \; \; \; \; \; \; \; \;
  \; \; \; \; \; \; \; \; \; \; \; \;
 \omega^5_{\; \; 7} =  -{1\over R}\;  e^7, \\
 \omega^6_{\; \; 7}& = & -{1\over R} cot\Theta \;  e^7
 -{1\over r} \; e^{-\Psi } \; cot\Theta \; (\hat{r} \cdot \hat{\Theta}) \;  e^0
 -{1\over r} e^{-\Psi } \;  (\hat{r} \cdot \hat{R}) \; e^0   \nonumber \\
 & & +{1\over r}  cot\Theta \; (\hat{\Phi} \cdot \hat{\Theta}) \;  e^2
      +{1\over r} (\hat{\Phi} \cdot \hat{R}) \; e^2
      -{1\over r}  cot\Theta \; (\hat{\theta} \cdot \hat{\Theta}) \;  e^3
      -{1\over r}   (\hat{\theta} \cdot \hat{R}) \;  e^3.
\end{eqnarray}
The  nonzero components of the curvature tensor are
\begin{eqnarray}
\bar{R}_{0101} &=& e^{-2\Lambda} ( \Psi^{''} -\Psi^{'} \Lambda^{'}
+ (\Psi^{'} )^2  )  -{3\over 4}{R^2 \over r^4}\;  e^{-2\Psi -2\Lambda}
\{ (\hat{r} \cdot \hat{\Theta})^2 +  (\hat{r} \cdot \hat{\Phi})^2 \},
 \\
\bar{R}_{0116}  &=& \; \; {R\over 2r^2}\;  e^{-\Psi -2\Lambda} (\Psi^{'} +
\Lambda^{'} + {2\over r})
(\hat{r} \cdot \hat{\Phi}), \\
\bar{R}_{0117}  &=& - {R\over 2r^2} \; e^{-\Psi -2\Lambda} (\Psi^{'} +
\Lambda^{'} + {2\over r})
(\hat{r} \cdot \hat{\Theta}),\\
\bar{R}_{0123}  &=& \; \; {R^2 \over 2r^4}\;  e^{-\Psi -\Lambda}
\{ (\hat{r} \cdot \hat{\Theta})^2 +  (\hat{r} \cdot \hat{\Phi})^2 \},\\
\bar{R}_{0156}  &=& - {1\over r^2} e^{-\Psi -\Lambda} (\hat{r} \cdot \hat{\Phi}),
\; \; \; \; \; \; \; \; \; \; \; \;
\bar{R}_{0157}  = \; \;{1\over r^2} e^{-\Psi -\Lambda} (\hat{r} \cdot \hat{\Theta}),\\
\end{eqnarray}
\begin{eqnarray}
\bar{R}_{0167}  &=&  - {1\over r^2} e^{-\Psi -\Lambda} (\hat{r} \cdot \hat{R}),
\; \; \; \; \; \;\; \; \; \; \; \;\; \; \; \; \;
\bar{R}_{0202} = \; \; {1\over r}\; \Psi^{'} e^{-2\Lambda},\\
\bar{R}_{0213}  &= & \; \; {R^2 \over 4r^4}\;  e^{-\Psi -\Lambda}
\{ (\hat{r} \cdot \hat{\Theta})^2 +  (\hat{r} \cdot \hat{\Phi})^2 \}, \\
\bar{R}_{0226} &=& - {R \over 2r^3}\;  e^{-\Psi -2\Lambda}
 (\hat{r} \cdot \hat{\Phi}) ,
 \; \; \; \; \; \;\; \;\; \; \; \; \;
\bar{R}_{0227} =\; \;   {R \over 2r^3}\;  e^{-\Psi -2\Lambda}
 (\hat{r} \cdot \hat{\Theta}) ,\\
 \bar{R}_{0303}& =&\; \;{1\over r}\; \Psi^{'} e^{-2\Lambda},\\
 \bar{R}_{0312} &=& - {R^2 \over 4r^4}\;  e^{-\Psi -\Lambda}
\{ (\hat{r} \cdot \hat{\Theta})^2 +  (\hat{r} \cdot \hat{\Phi})^2 \},\\
\bar{R}_{0336} &=& -{R \over 2r^3}\;  e^{-\Psi -2\Lambda}
 (\hat{r} \cdot \hat{\Phi}),
 \; \; \; \; \; \;\; \;  \; \; \; \; \; \;
 \bar{R}_{0337} =\; \;   {R \over 2r^3}\;  e^{-\Psi -2\Lambda}
 (\hat{r} \cdot \hat{\Theta}) ,\\
 \bar{R}_{0516} &=& - {1 \over 2r^2}\;  e^{-\Psi -\Lambda}
 (\hat{r} \cdot \hat{\Phi}) ,
 \; \; \; \; \; \;\; \; \; \; \; \;\; \; \; \;
 \bar{R}_{0517} =\; \;   {1 \over 2r^2}\;  e^{-\Psi -\Lambda}
 (\hat{r} \cdot \hat{\Theta}) ,\\
 \bar{R}_{0606} &=& \; \;  {R^2 \over 4r^4}\;  e^{-2\Psi -2\Lambda}
 (\hat{r} \cdot \hat{\Phi})^2 ,
 \; \; \; \; \; \;\; \;\; \; \; \; \;
 \bar{R}_{0607} = -  {R^2 \over 4r^4}\;  e^{-2\Psi -2\Lambda}
 (\hat{r} \cdot \hat{\Theta}) (\hat{r} \cdot \hat{\Phi}),\\
 \bar{R}_{0615} &=& \; \;  {1 \over 2r^2}\;  e^{-\Psi -\Lambda}
 (\hat{r} \cdot \hat{\Phi}) ,
 \; \; \; \; \; \;\; \; \; \; \; \;\;\; \; \; \;
\bar{R}_{0617}  = -  {1 \over 2r^2}\;  e^{-\Psi -\Lambda}
 (\hat{r} \cdot \hat{R}) ,\\
 \bar{R}_{0707} &=& \; \;  {R^2 \over 4r^4}\;  e^{-2\Psi -2\Lambda}
 (\hat{r} \cdot \hat{\Theta})^2 ,
 \; \; \; \; \; \;\; \; \; \; \; \; \;
 \bar{R}_{0715}  = -  {1 \over 2r^2}\;  e^{-\Psi -\Lambda}
 (\hat{r} \cdot \hat{\Theta}) ,\\
 \bar{R}_{0716} &=& \; \;   {1 \over 2r^2}\;  e^{-\Psi -\Lambda}
 (\hat{r} \cdot \hat{R}) ,
 \; \; \; \; \; \;\; \; \; \; \; \; \;\; \; \; \;
 \bar{R}_{1212} = \; \;{1\over r}\; \Lambda^{'} e^{-2\Lambda},\\
 \bar{R}_{1236} &=&\; \;   {R \over 2r^3}\;  e^{-\Lambda}
 (\hat{r} \cdot \hat{\Phi}) ,
 \; \; \; \; \; \;\; \; \; \; \; \; \; \; \; \; \;\; \; \; \;
 \bar{R}_{1237}  = -  {R \over 2r^3}\;  e^{-\Lambda}
 (\hat{r} \cdot \hat{\Theta}) ,\\
 \bar{R}_{1313} &=& \; \;{1\over r}\; \Lambda^{'} e^{-2\Lambda},
 \; \; \; \; \; \;\; \; \; \; \; \;\; \; \; \; \; \;\; \; \; \; \; \; \; \; \; \; \;
 \bar{R}_{1326}  = -  {R \over 2r^3}\;  e^{-\Lambda}
 (\hat{r} \cdot \hat{\Phi}) ,\\
 \bar{R}_{1327} &=& \; \;   {R \over 2r^3}\;  e^{-\Lambda}
 (\hat{r} \cdot \hat{\Theta}) ,
 \; \; \; \; \; \;\; \; \; \; \; \;\; \; \; \; \; \; \; \; \;
 \bar{R}_{1616}  =  -  {R^2 \over 4r^4}\;  e^{-2\Psi -2\Lambda}
 (\hat{r} \cdot \hat{\Phi})^2 ,\\
  \bar{R}_{1617} &=& \; \;   {R^2 \over 4r^4}\;  e^{-2\Psi -2\Lambda}
 (\hat{r} \cdot \hat{\Theta}) (\hat{r} \cdot \hat{\Phi}),
 \; \; \; \; \;
 \bar{R}_{1623}  = - {R \over r^3}\;  e^{ -\Lambda}
 (\hat{r} \cdot \hat{\Phi}) ,\\
 \bar{R}_{1717} &=& -  {R^2 \over 4r^4}\;  e^{-2\Psi -2\Lambda}
 (\hat{r} \cdot \hat{\Theta})^2 ,
 \; \; \; \; \; \; \; \; \; \; \; \; \,
 \bar{R}_{1723}  =   {R \over r^3}\;  e^{ -\Lambda}
 (\hat{r} \cdot \hat{\Theta}) ,\\
 \bar{R}_{2323} &=& \; \; {1\over r^2}(1-e^{-2\Lambda} )
 -{3\over 4}{R^2 \over r^4}\;
\{ (\hat{r} \cdot \hat{\Theta})^2 +  (\hat{r} \cdot \hat{\Phi})^2 \},\\
\bar{R}_{2356} &=& \; \;  {1 \over r^2}\;
 (\hat{r} \cdot \hat{\Phi}) ,
 \; \; \; \; \; \; \; \; \; \; \; \; \; \; \; \; \; \; \; \; \; \; \; \; \; \; \; \; \;
 \bar{R}_{2367}  =  \; \;  {1 \over r^2}\;
 (\hat{r} \cdot \hat{R}) ,\\
 \bar{R}_{2357} &=& -  {1 \over r^2}\;
 (\hat{r} \cdot \hat{\Theta}) ,
 \; \; \; \; \; \; \; \; \; \; \; \; \; \; \; \; \; \; \; \; \; \; \; \; \; \; \; \;
  \bar{R}_{2536}  =  \; \;  {1 \over 2r^2} \;
 (\hat{r} \cdot \hat{\Phi}) ,\\
  \bar{R}_{2537} &=& -  {1 \over 2r^2} \;
 (\hat{r} \cdot \hat{\Theta}) ,
 \; \; \; \; \; \; \; \; \; \; \; \; \; \; \; \; \; \; \; \; \; \; \; \; \; \; \,
 \bar{R}_{2626}  =  \; \; {R^2\over 4r^4}\; (\hat{r} \cdot \hat{\Phi})^2  \\
  \bar{R}_{2627} &=& -  {R^2 \over 4r^4}\;
 (\hat{r} \cdot \hat{\Theta}) (\hat{r} \cdot \hat{\Phi}),
  \; \; \; \; \; \; \; \; \; \; \; \; \; \; \; \; \;
 \bar{R}_{2635} = - {1 \over 2r^2}\;
 (\hat{r} \cdot \hat{\Phi}) \\
  \bar{R}_{2637} &=& \; \;  {1 \over 2r^2}\;
 (\hat{r} \cdot \hat{R}),
 \; \; \; \; \; \; \; \; \; \; \; \; \; \; \; \; \; \; \; \; \; \; \; \; \; \; \; \,
 \bar{R}_{2727}  =  \; \; {R^2\over 4r^4}\; (\hat{r} \cdot \hat{\Theta})^2  \\
 \bar{R}_{2735} &=& \; \;  {1 \over 2r^2}\;
 (\hat{r} \cdot \hat{\Theta}),
 \; \; \; \; \; \; \; \; \; \; \; \; \; \; \; \; \; \; \; \; \; \; \; \; \; \; \; \,
 \bar{R}_{2736} = -  {1 \over 2r^2}\;
 (\hat{r} \cdot \hat{R})\\
 \bar{R}_{3636}  &=&  \; \; {R^2\over 4r^4}\; (\hat{r} \cdot \hat{\Phi})^2 ,
 \; \; \; \; \; \; \; \; \; \; \; \; \; \; \; \; \; \; \; \; \; \; \; \; \; \; \,
  \bar{R}_{3637}  =  -  {R^2 \over 4r^4}\;
 (\hat{r} \cdot \hat{\Theta}) (\hat{r} \cdot \hat{\Phi}),\\
 \bar{R}_{3737}  &=&  \; \; {R^2\over 4r^4}\; (\hat{r} \cdot \hat{\Theta})^2 .
  \end{eqnarray}
We have tacitly omitted nearly one half of the nonzero components
 of the curvature tensor, because of the relations in (34) ,
 for shorthand.
 \section{Einstein Equation}
  Contracting the curvature tensor, one will get the Ricci tensor
 \begin{equation}
 \bar{R}_{\bar{a} \bar{b}} = \eta^{\bar{c} \bar{d}} \bar{R}_{\bar{c}\bar{a}\bar{d}\bar{b}}.
 \end{equation}
 Substituting the components of the dyonic metric,
 \begin{eqnarray}
 e^{2\Psi} &=&  1-{r_{s} \over r} , \\
e^{2\Lambda} &=&   (1-{r_{s} \over r })^{-1},
\end{eqnarray}
where $r_s$ is the Schwarzschild radius,
then  we can obtain almost $\bar{R}_{\bar{a} \bar{b}} $  are
 zero except
 \begin{eqnarray}
\bar{R}_{00}& =& -{1\over 2}  {R^2 \over r^4} \{ ( \hat{r} \cdot \hat{\Theta} )^2
+( \hat{r} \cdot \hat{\Phi} )^2 \}, \\
\bar{R}_{11}& =& +{1\over 2}  {R^2 \over r^4} \{ ( \hat{r} \cdot \hat{\Theta} )^2
+( \hat{r} \cdot \hat{\Phi} )^2 \} ,\\
\bar{R}_{22}& =& - {1\over 2}{R^2 \over { r^4}} \{ ( \hat{r} \cdot \hat{\Theta} )^2
+( \hat{r} \cdot \hat{\Phi} )^2 \}, \\
\bar{R}_{33}& =& -{1\over 2} {R^2   \over { r^4}} \{ ( \hat{r} \cdot \hat{\Theta} )^2
+( \hat{r} \cdot \hat{\Phi} )^2 \}.
\end{eqnarray}
 Then one has the Ricci scalar curvature,
\begin{eqnarray}
\bar{R} &=& \bar{\eta}^{\bar{a} \bar{b}} \bar{R}_{\bar{a} \bar{b}}  \\
&=& - \bar{R}_{00} + \bar{R}_{11}
+  \bar{R}_{22} + \bar{R}_{33} \\
&=& 0.
 \end{eqnarray}
From the fields $\widetilde{\mathcal{F}}_{\mu \nu}^{m}$ in
(22) and (23), the components of the  Ricci tensor, $(84) \sim (87)$, can be recast into the form,
\begin{equation}
\bar{R}_{\bar{a} \bar{b}} = -{1\over 2}E_{\bar{a}}^{\;\;  \mu} \; E_{\bar{b}}^{\;\;  \nu}\;
\bar{g}^{\alpha \beta}  \gamma_{mn}
\widetilde{\mathcal{F}}_{\mu \alpha}^{m}
\widetilde{\mathcal{F}}_{\nu \beta}^{n},
\end{equation}
or
\begin{equation}
\bar{R}_{\mu \nu} = -{1\over 2}\bar{g}^{\alpha \beta}  \gamma_{mn}
\widetilde{\mathcal{F}}_{\mu \alpha}^{m}
\widetilde{\mathcal{F}}_{\nu \beta}^{n}.
\end{equation}
Since the identity,
$  \gamma_{mn}
\widetilde{\mathcal{F}}_{\mu \nu}^{m}
\widetilde{\mathcal{F}}^{\mu \nu n} =0$, holds, the right-hand side of the equation (92) can be
identified as $8\pi$ times
the stress-energy tensor of the Yang-Mills field,
\begin{equation}
\bar{R}_{\mu \nu} = 8\pi \bar{T}_{\mu \nu}, \; \; \; \;
\bar{T}_{\mu \nu} ={-1\over 16 \pi}\bar{g}^{\alpha \beta}  \gamma_{mn}
\widetilde{\mathcal{F}}_{\mu \alpha}^{m}
\widetilde{\mathcal{F}}_{\nu \beta}^{n}.
\end{equation}
 Then the Einstein equation, $\bar{R}_{AB}-{1\over 2}
 \bar{g}_{AB} \bar{R} = 8\pi \bar{T}_{AB} $, is satisfied, where some components of $
 \bar{T}_{AB}$ are zero,
 $\bar{T}_{\mu m}=0$
 and $\bar{T}_{ m n}=0.$

The $ SO(3)$ KK dyonic metric satisfies the  Einstein equation in
 the seven-dimensional. The results from  two different methods are coincident.
 .

\end{document}